# Design of Two Dimensional Unipolar (Optical) Orthogonal Codes Through One Dimensional Unipolar (Optical) Orthogonal Codes

Ram Chandra Singh Chauhan, MIEEE., Yatindra Nath Singh, SMIEEE., Rachna Asthana, MIEEE.

*Abstract*— **In this paper, an algorithm for construction of multiple sets of two dimensional (2D) or matrix unipolar (optical) orthogonal codes has been proposed. Representations of these 2D codes in difference of positions representation (DoPR) have also been discussed along-with conventional weighted positions representation (WPR) of the code. This paper also proposes less complex methods for calculation of auto-correlation as well as cross-correlation constraints within set of matrix codes. The multiple sets of matrix codes provide flexibility for selection of optical orthogonal codes' set in wavelength-hopping time-spreading (WHTS) optical code division multiple access (CDMA) system.**

## I. Introduction

THE two dimensional unipolar orthogonal codes play an important role in terms of better performance (cardinality and spectral efficiency) than the one dimensional unipolar orthogonal codes [3]. When one dimensional unipolar orthogonal codes [1,2] are used in Optical CDMA system, the dimension can be temporal, spectral or spatial. The temporally, spectrally or spatially placed optical pulses at the position of bit '1's from the code sequence. When the two dimensional orthogonal codes are used, any two dimensions can be considered simultaneously e.g., temporal – spectral (wavelength), temporal – spatial, or spatial – spectral. Two dimensional unipolar orthogonal codes are also called matrix orthogonal codes. The matrix codes with one of the dimensions as time. In this paper, we are focusing only on these. The optical pulses are placed at the position of bit '1's of the orthogonal code in the two dimensional plane. The two dimensional or matrix orthogonal codes can be defined with the help of $(L \times N)$ array of the pulses, representing bit '1' by presence of a pulse and '0' by absence of it. only along time axis as the code word synchronization. The codes in a set are characterized by constant weight w, maximum autocorrelation constraint $\lambda_a$ and maximum cross-correlation constraint $\lambda_c$ respectively. It may be noted that auto-correlation and cross-correlation are meaningful only along time axis as the code word synchronization problem can happen only along time axis. Let the matrix codes X and Y belong to the same code set $C(L \times N, w, \lambda_a, \lambda_c)$ and follow the autocorrelation and cross correlation properties [4]. Here the codes with $1 \leq (\lambda_a, \lambda_c) \leq w-1,$ are also called pseudo orthogonal codes.

For $\lambda = \lambda_a = \lambda_c$ , the Johnson's bound A deriving maximum code set size $Z(L \times N, w, \lambda)$ is given [5] by

$$Z(L \times N, w, \lambda) \leq \left\lfloor \frac{L}{w} \left\lfloor \frac{LN-1}{w-1} \cdots \left\lfloor \frac{LN-\lambda}{w-\lambda} \right\rfloor \right\rfloor \right\rfloor = J_A(L \times N, w, \lambda);$$

Here, $\lambda$ is the Maximum Collision Parameter (MCP). It gives the number of maximum collisions of bit '1' elements in the array between any two code words.

Many schemes have been proposed for the generation of 2-D OOCs [5 – 13]. These schemes are capable of giving one or more than one set of matrix orthogonal codes with the size of each set less than the Johnson bounds. In this paper, a simple algorithm has been proposed to search the multiple sets of 2D OOC while achieving the upper bound in this process. We have used the already generated 1D OOC [1] for improving the efficiency.

## II. Representation of Matrix Orthogonal Codes

A matrix orthogonal codes is a matrix $(L \times N)$ of binary elements (0,1) with weight $w$, i.e. total number of bit 1's in the matrix are $w$. It is known that a matrix orthogonal codes is a matrix $(L \times N)$ of binary elements (0,1) in each row and column with weight $w$, i.e. total number of bit 1's in the matrix are $w$. Since each column in a 2-D code represent a time slot, column-wise circular shifting of a 2-D unipolar code is considered to be still the same code. The matrix codes can be

Manuscript received August 12, 2013. This work was supported in part by the G.B. Technical University, Lucknow under Teacher-fellowship scheme.

R.C.S Chauhan, is with the Electronics and Communication Engineering Department, Pranveer Singh Institute of Technology, Kanpur, India, as Associate Professor. (e-mail: ram.hbti123@ gmail.com).

Y. N. Singh, is with the Department of Electrical Engineering, Indian Institute of Technology, Kanpur, India, as Professor (e-mail: ynsingh@iitk.ac.in).

R. Asthana, is with the Electronics Engineering Department, H.B.T.I. Kanpur, India, as Associate Professor (e-mail: rachnaasthana@rediffmail.com).

represented by the position of weighted bits. Each bit position is indicated by $a'b$. where '$a$' is column number and '$b$' is row number of weighted bit. Here, $1 \leq a \leq L$, $0 \leq b \leq N-1$.

**Example I:** For L=4, N=5, weight $w = 4$, suppose the code is
$\begin{bmatrix} 1 & 0 & 0 & 0 & 0 \\ 0 & 0 & 1 & 0 & 0 \\ 1 & 0 & 0 & 0 & 0 \\ 0 & 1 & 0 & 0 & 0 \end{bmatrix}$. Its weighted position representation (WPR) will be (1'0, 3'0, 4'1, 2'2,). We scan the bits by one column at a time, from top to bottom. The columns are chosen from left to right.

There are N representations for same matrix orthogonal code in weighted positions representation (WPR). One for every column-wise circular shifted version. These are as follows:
WPR (1'0, 3'0, 4'1, 2'2), WPR (4'0, 2'1, 1'4, 3'4),
WPR (2'0, 1'3, 3'3, 4'4), WPR (1'2, 3'2, 4'3, 2'4),
WPR (1'1, 3'1, 4'2, 2'3).

In a matrix orthogonal code, the difference of positions (DoP) of consecutive weighted columns remain same on every column-wise circular shifted version of the code.

The matrix code can be represented with difference of positions representation (DoPR).

In DoPR of matrix orthogonal code, the position of weighted bit is represented by (a'd), where '$a$' is row number of weighted bit and '$d$' is consecutive difference of column position of next weighted bit with position of column of the current weighted bit, here $1 \leq a \leq L$, $0 \leq d \leq N-1$. We scanning the bits by one column at a time, from top to bottom. The columns are chosen from left to right. It provides an unique representation of code irrespective of amount of circular shifting along the row.

**Example II:** Let the matrix orthogonal code X with L=4, N=5, weight $w= 7$, be

Code X = $\begin{bmatrix} 1 & 0 & 0 & 0 & 1 \\ 0 & 1 & 0 & 0 & 0 \\ 1 & 0 & 0 & 0 & 1 \\ 0 & 1 & 0 & 0 & 1 \end{bmatrix}$.

X (WPR) = (1'0, 3'0, 2'1, 4'1, 1'4, 3'4, 4'4);
X (DoPR) = (1'0, 3'1, 2'0, 4'3, 1'0, 3'0, 4'1).

There are N=5 columns. Five WPR of this code are possible given along with the DoPR of each version.

$\begin{bmatrix} 1 & 0 & 0 & 0 & 1 \\ 0 & 1 & 0 & 0 & 0 \\ 1 & 0 & 0 & 0 & 1 \\ 0 & 1 & 0 & 0 & 1 \end{bmatrix}, \begin{bmatrix} \text{WPR (1'0, 3'0, 2'1, 4'1, 1'4, 3'4, 4'4),} \\ \text{DoPR (1'0, 3'1, 2'0, 4'3, 1'0, 3'0, 4'1)} \end{bmatrix}$

$\begin{bmatrix} 0 & 0 & 0 & 1 & 1 \\ 1 & 0 & 0 & 0 & 0 \\ 0 & 0 & 0 & 1 & 1 \\ 1 & 0 & 0 & 1 & 0 \end{bmatrix}, \begin{bmatrix} \text{WPR (2'0, 4'0, 1'3, 3'3, 4'3, 1'4, 3'4),} \\ \text{DoPR (2'0, 4'3, 1'0, 3'0, 4'1, 1'0, 3'1)} \end{bmatrix}$

$\begin{bmatrix} 0 & 0 & 1 & 1 & 0 \\ 0 & 0 & 0 & 0 & 1 \\ 0 & 0 & 1 & 1 & 0 \\ 0 & 0 & 1 & 0 & 1 \end{bmatrix}, \begin{bmatrix} \text{WPR (1'2, 3'2, 4'2, 1'3, 3'3, 2'4, 4'4),} \\ \text{DoPR (1'0, 3'0, 4'1, 1'0, 3'1, 2'0, 4'3)} \end{bmatrix}$

$\begin{bmatrix} 0 & 1 & 1 & 0 & 0 \\ 0 & 0 & 0 & 1 & 0 \\ 0 & 1 & 1 & 0 & 0 \\ 0 & 1 & 0 & 1 & 0 \end{bmatrix}, \begin{bmatrix} \text{WPR (1'1, 3'1, 4'1, 1'2, 3'2, 2'3, 4'3),} \\ \text{DoPR (1'0, 3'0, 4'1, 1'0, 3'1, 2'0, 4'3)} \end{bmatrix}$

$\begin{bmatrix} 1 & 1 & 0 & 0 & 0 \\ 0 & 0 & 1 & 0 & 0 \\ 1 & 1 & 0 & 0 & 0 \\ 1 & 0 & 1 & 0 & 0 \end{bmatrix}, \begin{bmatrix} \text{WPR (1'0, 3'0, 4'0, 1'1, 3'1, 2'2, 4'2),} \\ \text{DoPR (1'0, 3'0, 4'1, 1'0, 3'1, 2'0, 4'3)} \end{bmatrix}$.

□

One can observe that in this example for every column wise circular shifting of the code, WPR of code changes but DoPR remain same. Only circular shifting of DoPR (1'0, 3'0, 4'1, 1'0, 3'1, 2'0, 4'3) without changing the numerical values. Thus DOPR can be used as a unique representation of two dimensional unipolar (optical) orthogonal codes.

**Lemma 1**: In DoPR of matrix orthogonal code $(a_1'd_1, a_2'd_2, ..., a_w'd_w)$, $d_1 + d_2 + ... + d_w = N$, where N is number of columns in the binary matrix orthogonal code. □

The DoPR of matrix orthogonal code $(a_1'd_1, a_2'd_2, ..., a_w'd_w)$ may be converted to WPR $(a_1'b_1, a_2'b_2, ..., a_w'b_w)$ and vice versa with $0^{th}$ column containing at least one bit '1' necessarily as follows using modulo N addition.

$b_1 = 0;$
$b_2 = b_1 + d_1;$
$b_3 = b_2 + d_2;$
$---$
$b_w = b_{w-1} + d_{w-1};$

## III. CALCULATION OF CORRELATION CONSTRAINTS OF MATRIX ORTHOGONAL CODES

The auto-correlation constraint of a matrix orthogonal code is the maximum number of overlapping bit 1's of matrix code with its one of the non-zero column-wise circular shifted versions [4],[5].

Let us take the matrix codes X and Y from same set with code parameter $(L \times N, w, \lambda_a, \lambda_c)$. The auto-correlation constraint is defined as maximum value of the following for all valid value of $\tau$.

$$\sum_{i=0}^{L-1} \sum_{j=0}^{N-1} x_{i,j} x_{i,j \oplus \tau}, \quad for \quad 0 < \tau \leq N-1$$

The cross-correlation constraint for a pair of matrix orthogonal codes is the maximum number of overlapping bit 1's between two different matrix codes or thus column-wise circular shifted versions [4,5].

The cross-correlation constraint for the pair of codes X and Y can be defined as maximum value of the following for all valid values of $\tau$.

$$\sum_{i=0}^{L-1}\sum_{j=0}^{N-1} x_{i,j} y_{i,j\oplus\tau}, \quad for \quad 0 \leq \tau \leq N-1.$$

**Lemma 3.3:** The auto-correlation constraint $\lambda_a$ for a set of matrix orthogonal codes is the maximum of auto-correlation constraints for all the codes in the set. Thus for each code X,

$$\lambda_a \geq (X) \cap (X_P), \quad (0 < p \leq N-1)$$

Here $(X)$ represent the WPR of matrix code X and $(X_P)$ represent the WPR of p times column wise right circular shifted version of code X. The $A \cap B$ represents the number of entries which are common in set A and B.

**Example III:** Suppose matrix code X, with its WPR (X) be

$$X = \begin{bmatrix} 1 & 0 & 0 & 0 & 1 \\ 0 & 1 & 0 & 0 & 0 \\ 1 & 0 & 0 & 0 & 1 \\ 0 & 1 & 0 & 0 & 1 \end{bmatrix}, (X) = WPR(1'0, 3'0, 2'1, 4'1, 1'4, 3'4, 4'4),$$

$(X_1)$ = WPR (1'0, 3'0, 4'0, 1'1, 3'1, 2'2, 4'2),

$(X_2)$ = WPR (1'1, 3'1, 4'1, 1'2, 3'2, 2'3, 4'3),

$(X_3)$ = WPR (1'2, 3'2, 4'2, 1'3, 3'3, 2'4, 4'4),

$(X_4)$ = WPR(2'0, 4'0, 1'3, 3'3, 4'3, 1'4, 3'4),

$(X) \cap (X_1) = 2$

$(X) \cap (X_2) = 1$

$(X) \cap (X_3) = 1$

$(X) \cap (X_4) = 2$

Hence the auto-correlation constraint for the code X is $\lambda_a = 2$.

Similarly the cross-correlation constraint for a pair of matrix orthogonal code is the maximum number of overlapping bit 1's of a matrix code weighted positions of one matrix code with the any of 'N' cyclic shifted representations in WPR of the other matrix code.

$\lambda_c \geq (X) \cap (Y_P), \quad \forall p \ (0 \leq p \leq N-1)$

Also

$\lambda_c \geq (Y) \cap (X_P), \quad \forall p \ (0 \leq p \leq N-1)$

Where $(X)$ represent to WPR of matrix code X and $(Y_P)$ represent to WPR of p times column wise circular shifted version of code Y and vice versa.

**Example IV:** Suppose matrix codes X and Y, with their WPR (X) and (Y) respectively be as follows.

$$X = \begin{bmatrix} 1 & 0 & 0 & 0 & 1 \\ 0 & 1 & 0 & 0 & 0 \\ 1 & 0 & 0 & 0 & 1 \\ 0 & 1 & 0 & 0 & 1 \end{bmatrix}, (X) = WPR(1'0, 3'0, 2'1, 4'1, 1'4, 3'4, 4'4),$$

for this code $\lambda_a = 2$.

$$Y = \begin{bmatrix} 1 & 0 & 0 & 0 & 1 \\ 1 & 0 & 1 & 0 & 0 \\ 0 & 0 & 1 & 0 & 0 \\ 0 & 1 & 0 & 0 & 1 \end{bmatrix}, (Y) = WPR (1'0, 2'0, 4'1, 2'2, 3'2, 1'4, 4'4),$$

for this code, one can verify that $\lambda_a = 2$.

$Y_1$ = WPR (1'0, 4'0, 1'1, 2'1, 4'2, 2'3, 3'3),

$Y_2$ = WPR (1'1, 4'1, 1'2, 2'2, 4'3, 2'4, 3'4),

$Y_3$ = WPR (2'0, 3'0, 1'2, 4'2, 1'3, 2'3, 4'4),

$Y_4$ = WPR (4'0, 2'1, 3'1, 1'3, 4'3, 1'4, 2'4).

$(X) \cap (Y) = 4$

$(X) \cap (Y_1) = 2$

$(X) \cap (Y_2) = 2$

$(X_P) \cap (3 + X_P) = 2$

$(X_P) \cap (4 + X_P) = 2$

Hence the cross-correlation constraint for the pair of codes X and Y will be $\lambda_c = 4$.

IV. FORMATION OF TWO DIMENSIONAL UNIPOLAR (OPTICAL) ORTHOGONAL CODES' SETS

*Step 1:* One dimensional unipolar (optical) orthogonal codes of length $n = LN$, weight 'w' with the maximum auto-correlation and cross-correlation constraint to be equal to '$w$-1' are identified using the procedure given in proposed in [1],[2]. Suppose DoPR $(a_1, a_2, ..., a_w)$ is an one dimensional code with code length $n = LN$ and code weight 'w' with $(a_1 + a_2 + ... + a_w = LN)$ and $a_w \geq (a_1, a_2, ..., a_{w-1})$. The set of one dimensional codes with maximum auto-correlation and cross-correlation constraint to be less than or equal to '$w$-1', can be obtained by varying $(a_1, a_2, ..., a_{w-1})$ in the range 1 to $a_w$ such that $a_w = LN - (a_1 + a_2 + ... + a_{w-1})$.

*Step 2:* The conversion of one dimensional codes into the corresponding two dimensional (matrix) codes as follows.

(i) The one dimensional codes (in DoPR) is converted to corresponding WPR representations [1,2].

(ii) The (WPR) form of 1-D code is converted into two dimensional code WPR form by dividing each weighted position by 'L' to get quotient 'b' and remainder 'a'. Here each a'b represent to each weighted position in matrix orthogonal code. Here 'a' stands for row position and 'b' stands for column position.

The matrix orthogonal code with a'b weighted positions can be converted into corresponding binary matrix orthogonal code. This binary matrix orthogonal code can be used to generate 'L' binary matrix orthogonal codes by every row wise circular shifting of the code. These 'L' binary matrix orthogonal codes in WPR are [(a'b), (a'(b+1)),(a'(b+2)),...,(a'(b+L-1))], $\forall a,b$.

(iii) Conversion of two dimensional code (WPR) as obtained in (ii) into two dimensional code (DoPR) by getting difference 'd' of two columns of consecutive bit 1's in circular order so that each weighted position is represented by (a'd) as in example III. It will be standard DoPR of two dimensional codes if it is obtained from one dimensional code represented in standard DoPR.

The two dimensional unipolar (optical) orthogonal code (in DoPR) can be converted into binary matrix orthogonal code by converting it to two dimensional unipolar (optical) orthogonal code (in WPR).

*Step 3:* For every matrix orthogonal code formed after step 2, auto-correlation constraint of the every code can be calculated as in example III. Only the codes with desired auto-correlation constraint are selected.

*Step 4:* Using maximal clique [20] search methods [15-19] among the selected codes, set of 2D orthogonal codes are formed with desired cross-correlation and auto-correlation constraints.

In all the proposed schemes [6-14], the two-dimensional or matrix codes and their sets are very specific for matrix dimension $(L \times N)$, weight 'w' and correlation constraints ($\lambda_a$ and $\lambda_c$). While in the proposed algorithm we can generate the sets of matrix pseudo orthogonal codes with maximum code size. These code set are designed for general values of matrix dimension $(L \times N)$, weight 'w' and correlation constraints ($\lambda_a$ and $\lambda_c$).

## V. RESULTS AND CONCLUSION

The proposed algorithm is a general algorithm to generate all possible uni-polar two-dimensional or matrix pseudo orthogonal codes for a given matrix dimensions (L x N) and weight 'w'. This algorithm also gives the multiple sets of these matrix codes with maximum size given by Johnson bounds. Using the proposed method we have designed the sets of matrix orthogonal codes with number of row L=4, number of column N=3, weight of the code w=3, auto-correlation constraint and cross-correlation constraint less than equal to 2. The details of design process are given in appendix. This algorithm can be extended to design three dimensional as well as multi dimensional pseudo orthogonal codes as well as their multiple sets.

## APPENDIX

Input L =4, N=3, weight w=3, $(\lambda_a, \lambda_c) \leq 2$

1. One dimensional DoP code {1 1 10}. Corresponding one dimensional in WPR {1,2,3} and binary code ( 1 1 1 0 0 0 0 0 0 0 0 0). Corresponding Two dimensional DoP code [1'0 2'0 3'0], Two-dimensional binary code $\begin{bmatrix} 1 & O & O \\ 1 & O & O \\ 1 & O & O \\ O & O & O \end{bmatrix}$

2. One dimensional DoP code {1 2 9}. Corresponding one dimensional code in WPR {1,2,4} and binary code ( 1 1 0 1 0 0 0 0 0 0 0 0). Corresponding two-dimensional DoP code [1'0  2'0  4'0], Two dimensional binary code
$$\begin{bmatrix} 1 & O & O \\ 1 & O & O \\ O & O & O \\ 1 & O & O \end{bmatrix}$$

3. One dimensional DoP code {1 3 8}. Corresponding one dimensional code in WPR {1,2,5} and binary code ( 1 1 0 0 1 0 0 0 0 0 0 0). Corresponding two dimensional DoP code [1'0  2'0  1'1], Two dimensional binary code
$$\begin{bmatrix} 1 & 1 & O \\ 1 & O & O \\ O & O & O \\ O & O & O \end{bmatrix}$$

4. One dimensional DoP code {1 4 7}. Corresponding one dimensional code in WPR {1,2,6} and binary code ( 1 1 0 0 0 1 0 0 0 0 0 0). Corresponding two dimensional DoP code [1'0  2'0  2'1], Two dimensional binary code
$$\begin{bmatrix} 1 & O & O \\ 1 & 1 & O \\ O & O & O \\ O & O & O \end{bmatrix}$$

5. One dimensional DoP code {1 5 6}. Corresponding one dimensional code in WPR {1,2,7} and binary code ( 1 1 0 0 0 0 1 0 0 0 0 0). Corresponding two dimensional DoP code [1'0  2'0  3'1], Two dimensional binary code
$$\begin{bmatrix} 1 & O & O \\ 1 & O & O \\ O & 1 & O \\ O & O & O \end{bmatrix}$$

6. One dimensional DoP code {2 1 9}. Corresponding one dimensional code in WPR {1,3,4} and binary code ( 1 0 1 1 0 0 0 0 0 0 0 0). Corresponding two dimensional DoP code [1'0  3'0  4'0], Two dimensional binary code
$$\begin{bmatrix} 1 & O & O \\ O & O & O \\ 1 & O & O \\ 1 & O & O \end{bmatrix}$$

.
.
.

18. One dimensional DoP code {4 4 4}. Corresponding one dimensional code in WPR {1,5,9} and binary code ( 1 0 0 0 1 0 0 0 1 0 0 0 ). Corresponding Two dimensional DoP code [1'0  1'1  1'2], Two dimensional binary code
$$\begin{bmatrix} 1 & 1 & 1 \\ O & O & O \\ O & O & O \\ O & O & O \end{bmatrix}$$

19. One dimensional DoP code {5 1 6}. Corresponding one dimensional code in WPR {1,6,7} and binary code ( 1 0 0 0 0 1 1 0 0 0 0 0 ). Corresponding two dimensional DoP code [1'0  2'1  3'1], Two dimensional binary code
$$\begin{bmatrix} 1 & O & O \\ O & 1 & O \\ O & 1 & O \\ O & O & O \end{bmatrix}$$